# A Novel Implementation of Machine Learning for the Efficient, Explainable Diagnosis of COVID-19 from Chest CT

**Justin Liu**

**Abstract**

In a worldwide health crisis as severe as COVID-19, there has become a pressing need for rapid, reliable diagnostics. Currently, popular testing methods such as reverse transcription polymerase chain reaction (RT-PCR) can have high false negative rates. Consequently, COVID-19 patients are not accurately identified nor treated quickly enough to prevent transmission of the virus. However, the recent rise of medical CT data has presented promising avenues, since CT manifestations contain key characteristics indicative of COVID-19. This study aimed to take a novel approach in the machine learning-based detection of COVID-19 from chest CT scans. First, the dataset utilized in this study was derived from three major sources, comprising a total of 17,698 chest CT slices across 923 patient cases. Additionally, image preprocessing algorithms were developed to reduce noise by excluding irrelevant features. Transfer learning was also implemented with the EfficientNetB7 pre-trained model to provide a backbone architecture and save computational resources. Lastly, several explainability techniques were leveraged to qualitatively validate model performance by localizing infected regions and highlighting fine-grained pixel details. The proposed model attained an overall accuracy of 92.71% and a sensitivity of 95.79%. Explainability measures showed that the model correctly distinguished between relevant, critical features pertaining to COVID-19 chest CT images and normal controls. Deep learning frameworks provide efficient, human-interpretable COVID-19 diagnostics that could complement a radiologist's decision or serve as an alternative screening tool. Future endeavors could provide insight into infection severity, patient risk stratification, and more precise visualizations.

## 1 Introduction

COVID-19 has unfolded into a global pandemic. The novel virus has affected over 91 million people worldwide and claimed over 2 million lives [1]. Its effects have pushed the world to the brink of social and economic collapse: placing countries in turmoil, quarantining human civilization, and ravaging countless industries. Moreover, it has become difficult to provide the necessary treatment to all patients [2, 3, 4], and thus there is a pressing need for rapid diagnostics. Although typical symptoms of the coronavirus include fever, dry cough, muscle pain, shortness of breath, fatigue, and headache, in some scenarios the virus can be asymptomatic [5], posing a tremendous public health threat. COVID-19 has proven to be one of the most severe public health crises in the past hundred years [6].

An early diagnosis of COVID-19 is imperative for disease control and containment. Currently, popular testing methods such as reverse-transcription polymerase chain reaction (RT-PCR) have high specificity but relatively lower sensitivity. The limited supply and strict requirements for laboratory environments also delay the diagnosis of patients [7]. Consequently, through RT-PCR testing, COVID-19 patients are not accurately identified nor treated quickly enough to prevent transmission of the virus.

On the contrary, with the recent rise of COVID-19 data from chest CT scans, studies to investigate this underlying concern are enabled. Results have manifested chest CT imaging to be more reliable in diagnostics and thereby effective in disease containment [2, 8]. Fang, Zhang, et al. [9] found that RT-PCR tests revealed a sensitivity of 71%, in comparison to chest CT imaging that attained a sensitivity of 98% [10]. This raises the prospect of having an alternative diagnostic tool for the novel virus's clinical management.

Concurrently, artificial intelligence and machine learning have witnessed monumental growth in bridging the gap between the capabilities of man and machine. The agenda for this field is to equip computers with the necessary data so that machines can view the world as humans do, perceive it in a similar manner, and even use the knowledge for a multitude of tasks such as image classification. With the emergence of medical CT data, machine learning can shed light on an efficient, accessible, and accurate diagnosis of the novel coronavirus.

Computer vision with deep learning [11], a subfield of artificial intelligence, has advanced over time, primarily using one particular algorithm--a convolutional neural network, otherwise known as a ConvNet or CNN. A ConvNet, or an artificial network of neurons, analyzes imagery where the details that would otherwise be difficult to interpret with the human eye are recognized by the computer. ConvNets have been utilized in the past to diagnose diseases, thereby complementing other diagnostic techniques in the healthcare industry. For radiologists, analyzing medical imagery is a time-consuming, manual procedure, particularly when patient volume is substantial. Receiving RT-PCR test results can also take several days due to backlogs or other priorities in the lab [12], so the efficiency of deep learning-based techniques' makes for an appealing trait that can benefit clinicians.

Deep learning-based techniques for COVID-19 detection have emerged in a multitude of works [6, 7, 13, 14, 15, 16, 17, 18, 19, 20, 21], which have suggested the potential of artificial intelligence. However, several studies thus far have utilized customized models. This paper aimed to employ a novel pre-trained model for transfer learning [22], a machine learning method where previous knowledge is transferred, or applied to a new task. Additionally, a diverse, multinational dataset with chest CT scans from three distinct sources was assembled, in contrast to traditional works that restrict their consideration of data to one source, limiting their generalizability. Although investigations [7, 14, 15, 17, 18, 19, 20, 21, 23, 24] have surveyed transfer learning, they either raise the same concern in using data from a single source, lack a wide variety of human-interpretable measures, or choose to analyze X-rays. While X-rays are slightly more accessible than CT scans, CT scans offer a much higher level of detail that generates 360-degree views. CT imagery also provides a more useful understanding of soft tissue, blood vessels, and inflammation, all of which X-rays fail to show [25, 26]. Furthermore, CT manifestations contain key points indicative of COVID-19 such as ground-glass opacities, consolidation, reticular pattern, and crazy-paving patterns [27].

For transfer learning, the EfficientNetB7 pre-trained model [28] was employed, a backbone architecture that has surpassed state-of-the-art accuracies with up to 10 times better efficiency, attaining 84.4% top-1 and 97.1% top-5 accuracy on ImageNet [29]. The EfficientNetB7 architecture outperforms popular base models such as ResNet-50 [30] and Inception-v3 [31], presenting a promising alternative to conventional methods.

In addition to transfer learning, this study aims to take a unique approach in utilizing Gradient-weighted Class Activation Mapping (Grad-CAM) [32] for visualizable, human-interpretable examination. To the best of the author's knowledge, this paper is the first to leverage Guided Backpropagation [33] to localize the fine-grained pixel annotations that Grad-CAM lacks. Since machine learning is often a black box where one cannot visualize the intricacies within a model's decision-making process, Grad-CAM sheds light on explainability measures, diminishing the inexplicable nature of artificial intelligence. This in turn allows for the validation of relevant CT features considered by the model in its decision-making behavior.

The objective of this paper was to take a novel approach in assembling a generalizable deep learning framework that could diagnose COVID-19 from chest CT in an efficient yet explainable way. It was hypothesized that the use of several sources of data, transfer learning with a computationally efficient base model, and a wide variety of explainability techniques would help achieve such an objective. Real-world deployment on a web application is another goal, in which the deep learning framework would not replace, but instead complement RT-PCR or a radiologist's diagnosis.

## 2 Methods

### 2.1 Dataset

This study employed a dataset comprising a total of 17,698 chest CT scans across 923 patient cases. Data were derived from three major sources: the China National Center for Bioinformation (CNCB) [34], The Cancer Imaging Archive (TCIA) [35], and COVID-CTset [36].

The CNCB dataset is an open-source dataset of the lung CT images and metadata constructed from various cohorts from the China Consortium of Chest CT Image Investigation (CC-CCII). The images were classified into normal controls, common pneumonia, and novel coronavirus pneumonia (NCP) due to SARS-CoV-2, but since the volume of data surrounding common pneumonia was not large enough for the model to accurately learn from, common pneumonia was excluded. Despite the CNCB comprising over 100,000 images, computational resources did not permit such an immense amount of data to be handled. Hence, a total of 6,173 slices with lesions identified by CC-CCII were randomly selected. It was also ensured that volumes with segmented lung regions were excluded because this study aimed to examine chest CT images that are normally used in diagnostics.

Another source of data was derived from TCIA, which consisted of unenhanced chest CTs from 632 patients with positive RT-PCR for SARS-CoV-2 and ground truth annotations of COVID-19 lesions in the lung. The COVID-19 Lung CT Lesion Segmentation Challenge selected 249 patients from the TCIA dataset, from which an algorithm was further developed to select slices with visible lungs in the image sequences of a patient--resulting in a final 4,243 images.

COVID-CTset, the last source of data, contained the full, original 15,589 and 48,260 CT scans belonging to 95 COVID-19 positive and 282 normal persons, respectively. Data was gathered from Negin Medical Center, located in Sari, Iran. Again, due to a computational restraint, a total of 7,282 images from the dataset were randomly selected for use.

Finally, the assembled dataset used an approximate 80%-10%-10% split for training, validation, and testing data, respectively.

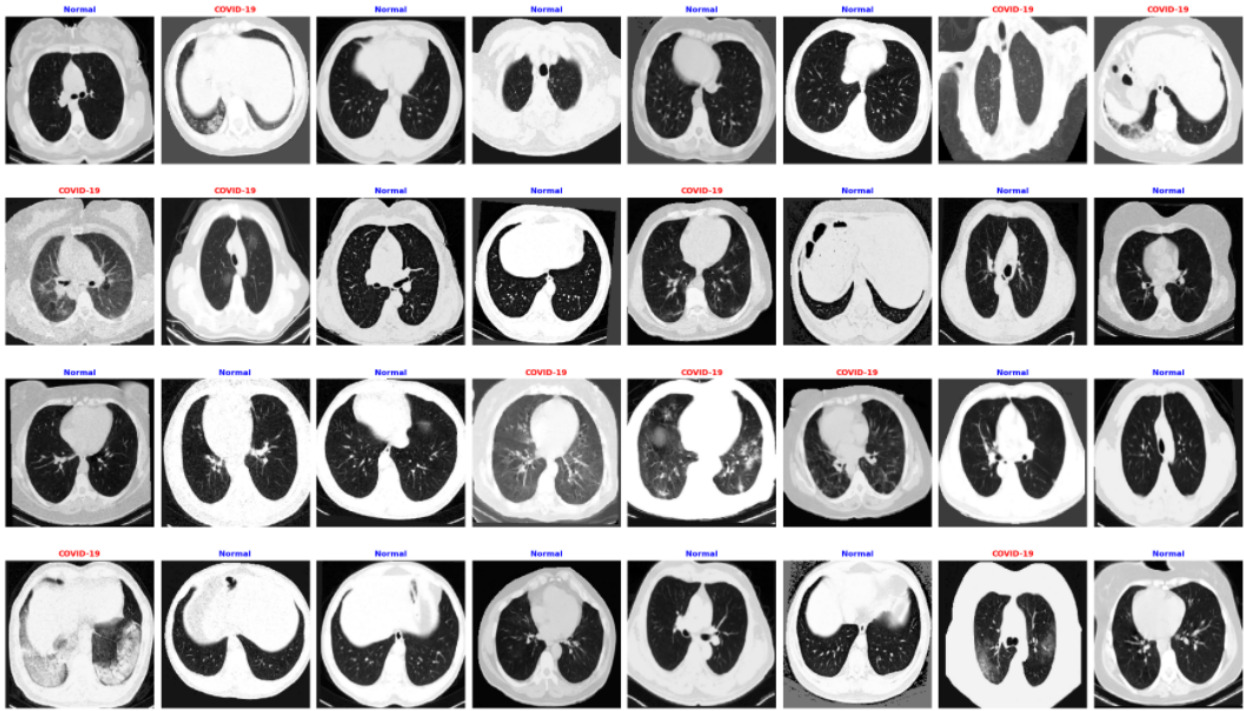

Figure 1: A sample batch of images from the dataset utilized in this study. Aggregating data from three distinct sources increased generalizability, but it also made the learning process a more rigorous task.

## 2.2 Data Preprocessing

To standardize all images that enter the network, an automatic body cropping method that involved a series of morphological transformations was developed. To begin the process, all images were converted to the same data type: 8-bit unsigned integer. This data type ensured that all pixel values within an image were normalized to range from 0 to 255. Next, the image was Gaussian smoothed to reduce image noise. A pixel threshold was then applied to the image, where values below the threshold were set to 0, and values above the threshold were set to 255, resulting in a binary image. From there, a set of morphological operations applied a structuring element to the images. The two operations utilized were erosion and dilation. Erosion first removed pixels resting along the edges of a binary region or blob. Dilation followed by expanding the remaining pixels around the edges of a binary region or blob. The effect of the erosion and dilation maneuver, known as binary opening, was that it erased small blobs and thin regions, which in a chest CT scan is often noise from the bed or imaging artifacts. At this point, the body mask of the original CT scan remained, with nearly all noise removed. The major contours are then determined by the computer, and the contour with the largest area was the bounding box to which the original image was cropped. Figure 2 depicts the automatic lung field crop at each stage of the procedure.

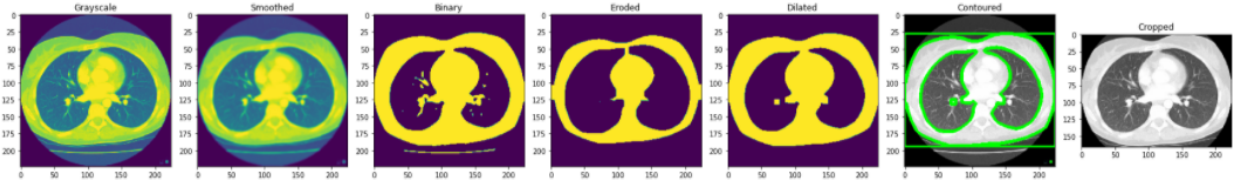
Figure 2: The stages of an automatic lung field crop on a sample chest CT scan. After ensuring the image has only one color channel (grayscale), smoothing the image out, and setting a pixel threshold to create a binary image, erosion and dilation remove irrelevant features. The image is then cropped to the bounding box created around the contour with the largest area.

While cropping images assists in directing the machine's focus to the lungs, noise contamination from the CT scanner bed and imaging artifacts are still present, which can adversely affect model performance. To solve this, another preprocessing algorithm was employed, as illustrated in Figure 3. It is also worth noting that this study did not include CT volumes where the entire background was removed to leave a segmented lung region, as the contrast between the segmented lungs and the background can lead to biases in a model's decision-making behavior. Excluding segmented CT volumes also enabled the model to grasp a better understanding of what CT scans typically look like--which are unsegmented.

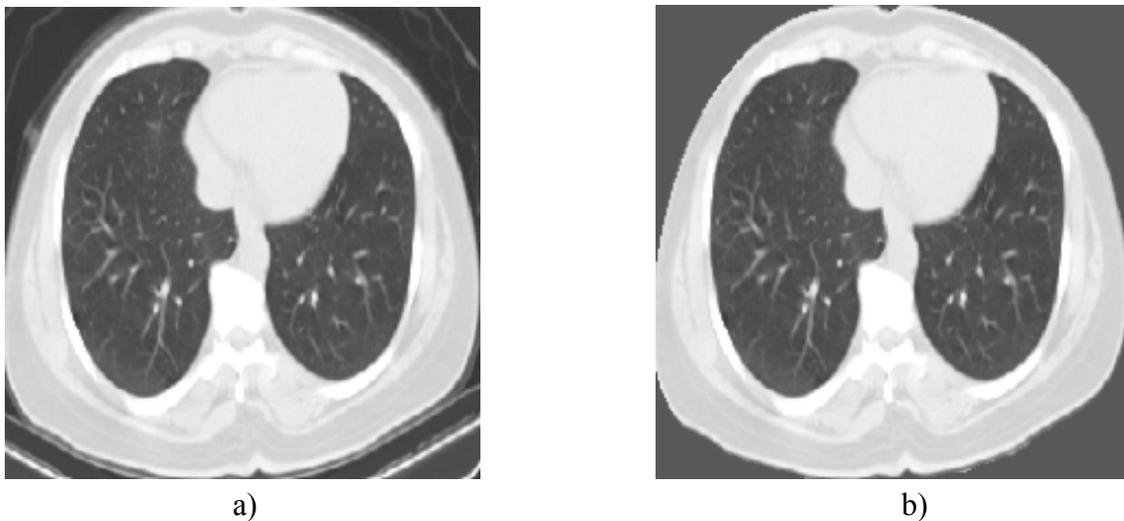

a)  b)

Figure 3: A chest CT scan before (a) and after (b) the replacement of noise from external features with a background mean.

As previously mentioned, the TCIA [35] dataset was employed in this study. However, unlike standard image formats that store files in Portable Network Graphics (PNG) format, the TCIA dataset stored their images in Neuroimaging Informatics Technology Initiative (NIfTI), a special file format for neuroimaging [37]. Thus, it was necessary to develop an algorithm to process and convert the NIfTI files into PNG images. To do so, the NiBabel Python library [38] was leveraged. All images within a patient's CT sequence were iterated over, but only slices near the middle of the sequence were used. These slices were indicative of "open" lungs in which characteristics of COVID-19 were visible. Additionally, all images were initially stored in 16-bit grayscale Hounsfield units [39], so the previously uninterpretable images were converted into standard 8-bit, grayscale images. Afterward, the file was cropped to remove exterior noise and saved as a PNG file. In regards to the COVID-CTset [36], all images were also stored in Hounsfield units, so the identical approach was used.

Although the dataset assembled in this study already varies in images from three sources, data augmentation was leveraged for further diversity. Data augmentation artificially created new data from existing data by randomly rotating, flipping, shearing, brightening, darkening, translating, and zooming in on images. This technique aided the model in generalizing better to unseen data.

Finally, before feeding the images into the ConvNet, all images were rescaled by dividing the pixel values by 255, which normalized the data to 32-bit floating-point format and range from 0 to 1. This was essential in simplifying the learning process and lowering computational resources. A target size of 224 x 224 pixels was also applied to each image because it provided a reasonable tradeoff between computation and image detail.

### 2.3 Transfer Learning and Network Architecture

The network architecture performed transfer learning with a base model pre-trained on ImageNet [29] coupled with Adam, an adaptive gradient descent optimizer. Generally, training a model from scratch for large datasets is computationally demanding and time-consuming. The pre-trained model with transfer learning enables the facility to speed up convergence, or the progression of error minimization, with network generalization [6]. In transfer learning, weights and biases are transferred from a pre-trained model, thereby providing a backbone that can detect basic patterns or edges. A number of popular pre-trained models have been put into practice: GoogleNet [40], LeNet [41], SqueezeNet [42], Xception [43], variations of VGG [44], Inception [45], MobileNets [46], DenseNet [47], U-Net [48], and different forms of ResNet [49]. However, in this study, a different approach was taken by leveraging the novel EfficientNetB7 model [28]. EfficientNetB7 was selected from a family of EfficientNets, where compound scaling of depth, width, and image resolution facilitated the ideal relationship between different dimensions under a fixed computational budget (FLOPs, or floating-point operations).

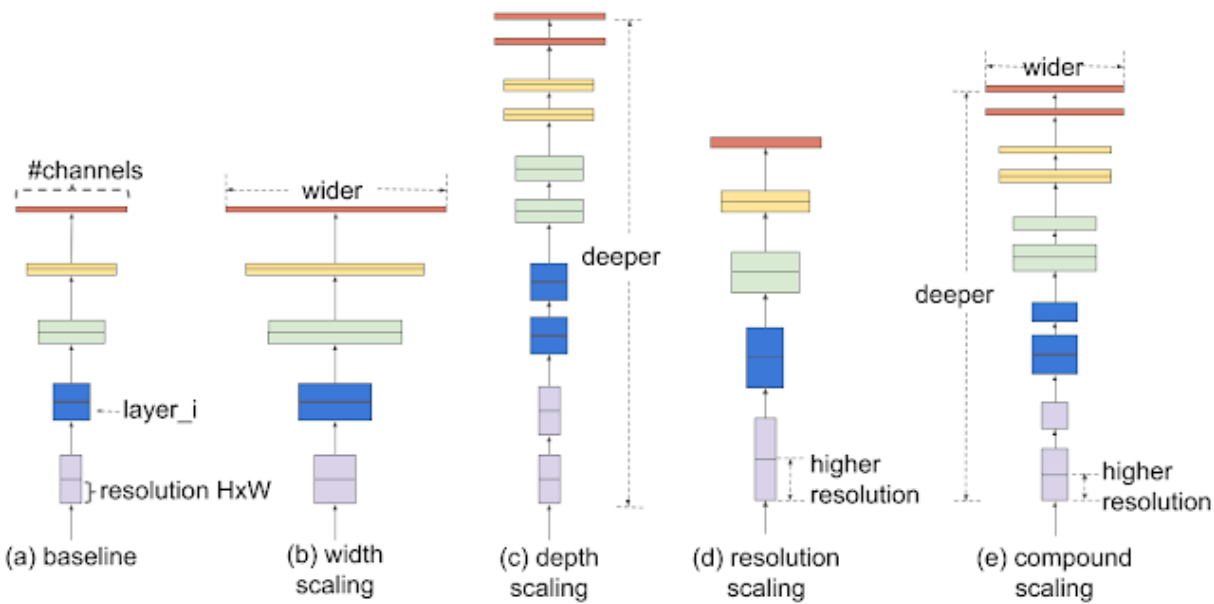

Figure 4: In contrast to conventional practices (b)-(d), EfficientNets' compound scaling method

(e) uniformly scales up all three dimensions: depth, width, and image resolution. *Credit: (Tan & Le, 2019)*

To implement transfer learning, previously existing layers were first frozen to avoid destroying any of the information they contained during future training rounds. Global average pooling followed to reduce spatial dimensions. Finally, a new fully connected layer with 512 neurons leading up to a final fully connected layer with a single output neuron was added to fit the purposes of this classification task.

**2.4 Grad-CAM Visualizations**

For explainability measures, Grad-CAM [32] was leveraged for visualizable, human-interpretable inspection. This was done by calculating the gradient of a predicted class with respect to the final convolutional layer's activations, and then weighting those activations with the calculated gradient. While the generated output of this procedure, a heatmap, does provide insight into prominent regions of importance, it fails to show specific details and features relevant to the model's decision. Thus, the novel use of Guided Backpropagation [33] to highlight fine-grained pixel annotations in CT imagery was employed in this study. Furthermore, to include the best from both worlds, a Guided Grad-CAM is calculated via an element-wise multiplication between Grad-CAM and Guided Backpropagation to include high-resolution and class-discriminative image regions. Because the decision-making process of a ConvNet is a black box, explainability analysis aids in a better understanding of model decision-making behavior and image features that were critical to the final prediction. Additionally, this phase of human interpretation allowed for verification that the model was making inferences from relevant factors, and not erroneous noise.

**3 Results**

The test set with 1,784 unseen images was used to evaluate model performance. The test set included 906 COVID-19 positive chest CT images aggregated from CNCB [34], TCIA [35], and COVID-CTset [36] with 878 normal controls from CNCB and COVID-CTset. This diverse set of data provided a better representation of the possible CT images the model could encounter in practice. In addition to analyzing quantitative values, explainability measures validated qualitative results through identifying factors critical to the model's decision-making behavior.

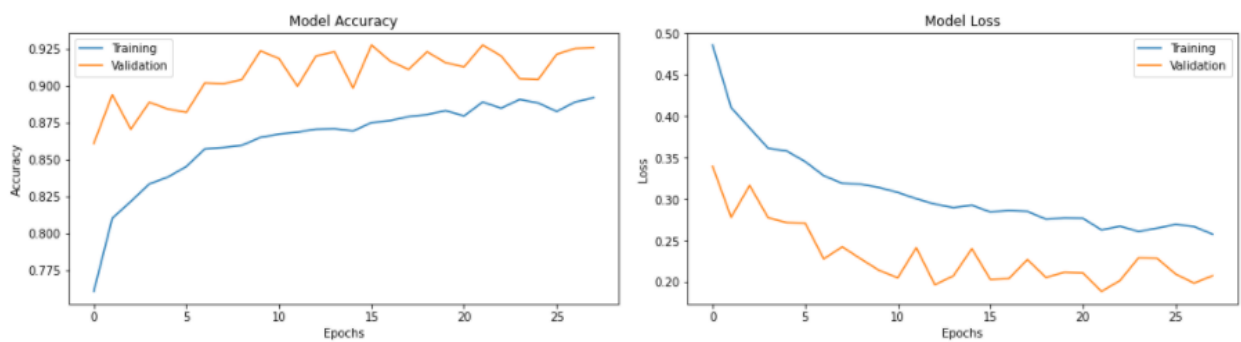

Figure 5: Training and validation plots of accuracy and loss are shown. The early stopping callback halted the learning process at 28 epochs to prevent overfitting.

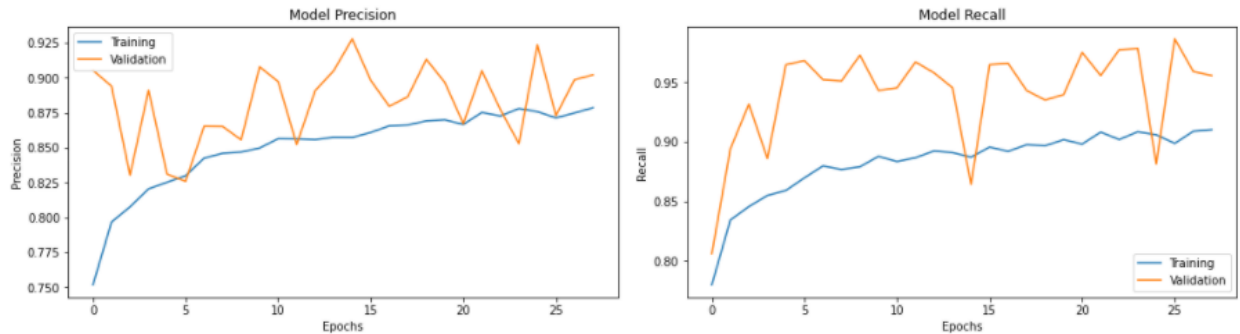

Figure 6: Training and validation plots of precision and recall are shown. The early stopping callback halted the learning process at 28 epochs to prevent overfitting.

As depicted in Figures 5 and 6, coupled with early stopping, which halted the learning process when metrics stopped improving, the proposed model was trained for 50 epochs, or the number of times all the training samples passed through the machine learning algorithm.

When evaluated on the test set, an overall accuracy of 92.71%, precision of 90.04%, recall of 95.79%, and F1-score of 92.83% were attained.

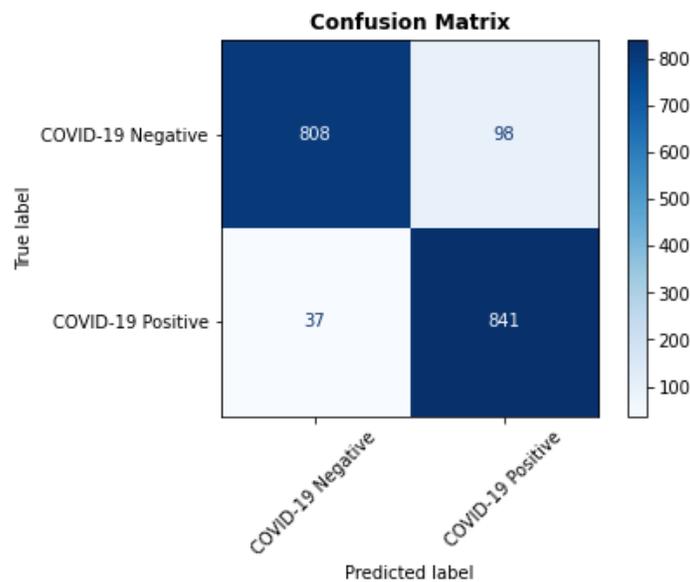

Figure 7: Binary classification confusion matrix for the proposed model trained with exterior exclusion.

The aforementioned metrics are observed in Figure 7, where the true negatives (upper-left) denotes the number of non-COVID-19 subjects who are correctly classified as not having the infection, false positives (upper-right) denotes the number of non-COVID-19 patients who are misidentified as having the infection, false negatives (bottom-left) denotes the number of COVID-19 subjects who are misclassified as healthy, and true positives (bottom-right) denotes the number of COVID-19 patients correctly identified by the model as having the infection.

Table 1: COVID-19 classification report for the best models trained with and without exterior exclusion.

| Training Method | Accuracy (%) | Precision (%) | Recall (%) | F1-score (%) |
|---|---|---|---|---|
| No Exterior Exclusion | 92.71 | 90.04 | 95.79 | 92.83 |
| Exterior Exclusion | 92.94 | 89.63 | 96.60 | 92.98 |

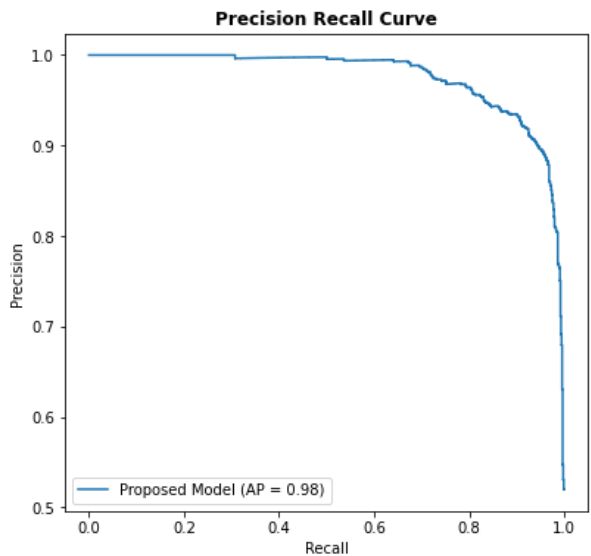

Figure 8: Precision-recall curve for the proposed model trained with exterior exclusion. Average precision was calculated to summarize the curve into one value that accommodates precision at every possible decision threshold.

The precision-recall curve plots the precision along the y-axis and the recall along the x-axis, providing a visual for the precision-recall tradeoff. An ideal model would have a coordinate point at (1, 1), and a skillful model would have a curve that approaches the coordinate point of (1, 1), as seen in Figure 8. In contrast, a classifier with no skill cannot discriminate between the classes, and its decisions would be based purely on random guessing. Thus, its plot would simply have a horizontal line at a precision proportional to the number of positive examples in the dataset.

While precision does provide valuable insights into model performance, it refers to a particular decision threshold. However, classes are not always balanced, and one may want to vary the decision threshold. Thus, the average precision was calculated to account for precision at all possible thresholds and summarize the entire precision-recall curve into a single value. In this study, an average precision of 98% was achieved.

The F1 score was calculated to find the harmonic mean of precision and recall. In contrast to simply taking the average of the two metrics, the harmonic mean penalizes extreme values. For example, if the precision was 0% and the recall was 100%, taking the average would return 50%. In contrast, the harmonic mean would be calculated as 0%. Therefore, the F1 score provides a balanced, comprehensive evaluation of model performance--all in a single percentage.

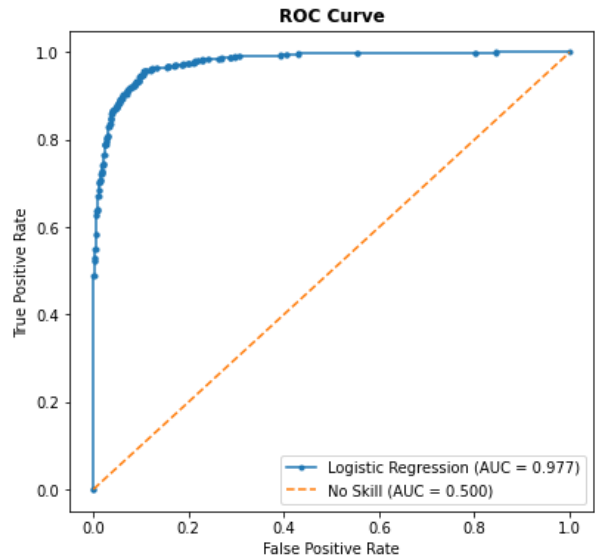

Figure 9: Receiver operator characteristic curve for the proposed model trained with exterior exclusion.

In addition to measuring precision, recall, and F1-score, the receiver operating characteristic curve, or ROC curve, and area under the curve, or AUC, were utilized. The ROC is created by plotting the true positive rate versus the false positive rate, and the AUC represents an overall performance of a classifier across all threshold settings. A classifier that has zero discriminative power between positive and negative classes will result in a diagonal line passing through the coordinate point (0,0) where both the false positive rate and the true positive rate are 0. In other words, the classifier's decision-making behavior is purely based on random guessing. Hence, because the ideal ROC curve has an AUC of 1 and the worst ROC curve has an AUC of 0.5, the AUC value of any useful binary classifier must range between 0.5 and 1. The proposed EfficientNetB7 transfer learning model obtained 0.977 ROC AUC.

Furthermore, the ROC curve illustrates a trade-off between the true positive rate and false positive rate, such that a change in the decision threshold will affect the true positive rate at the expense of the false positive rate, and vice versa. Consequently, the ROC curve can help determine the optimal threshold for the model's predictive behavior.

Aside from quantitative evaluation, qualitative analysis through various machine learning explainability techniques was leveraged. In the first row of both figures 10 and 11, the patient's CT scan is shown. The second row displays a superimposed image of the Grad-CAM heat map overlaid on the original image. Row three displays the Guided Backpropagation visualizations which highlight pixels that triggered greater neuron activations within the model. In contrast to Grad-CAM which can only localize general regions of interest, Guided Backpropagation can trace fine-grained details--a quality that, to the best of the author's knowledge, has not been utilized in any other work on the automated classification of COVID-19. Next, row four combines the best from Grad-CAM and Guided Backpropagation through an element-wise multiplication of the two, creating a Guided Grad-CAM. Lastly, row 5 presents a new explainability technique that was developed in this study. It resembles an enhanced version of Guided Grad-CAM, where the green color channel of Guided Grad-CAM was multiplied by 3, allowing for greater saliency and emphasis on key patterns.

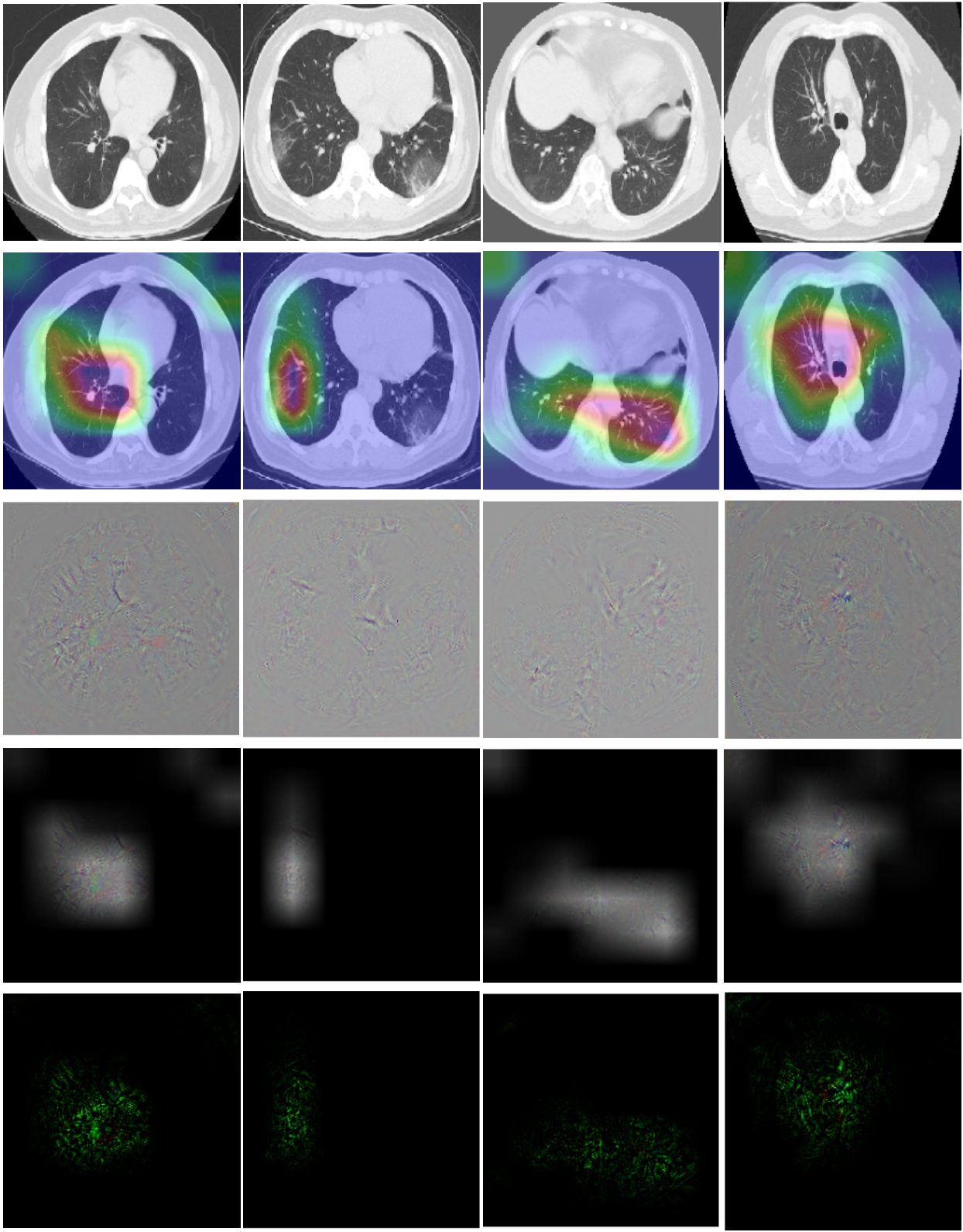

Figure 10: Sample chest CT images of COVID-19 cases from the test set and their associated critical features (e.g., ground-glass opacities and reticular pattern) identified by the proposed model.

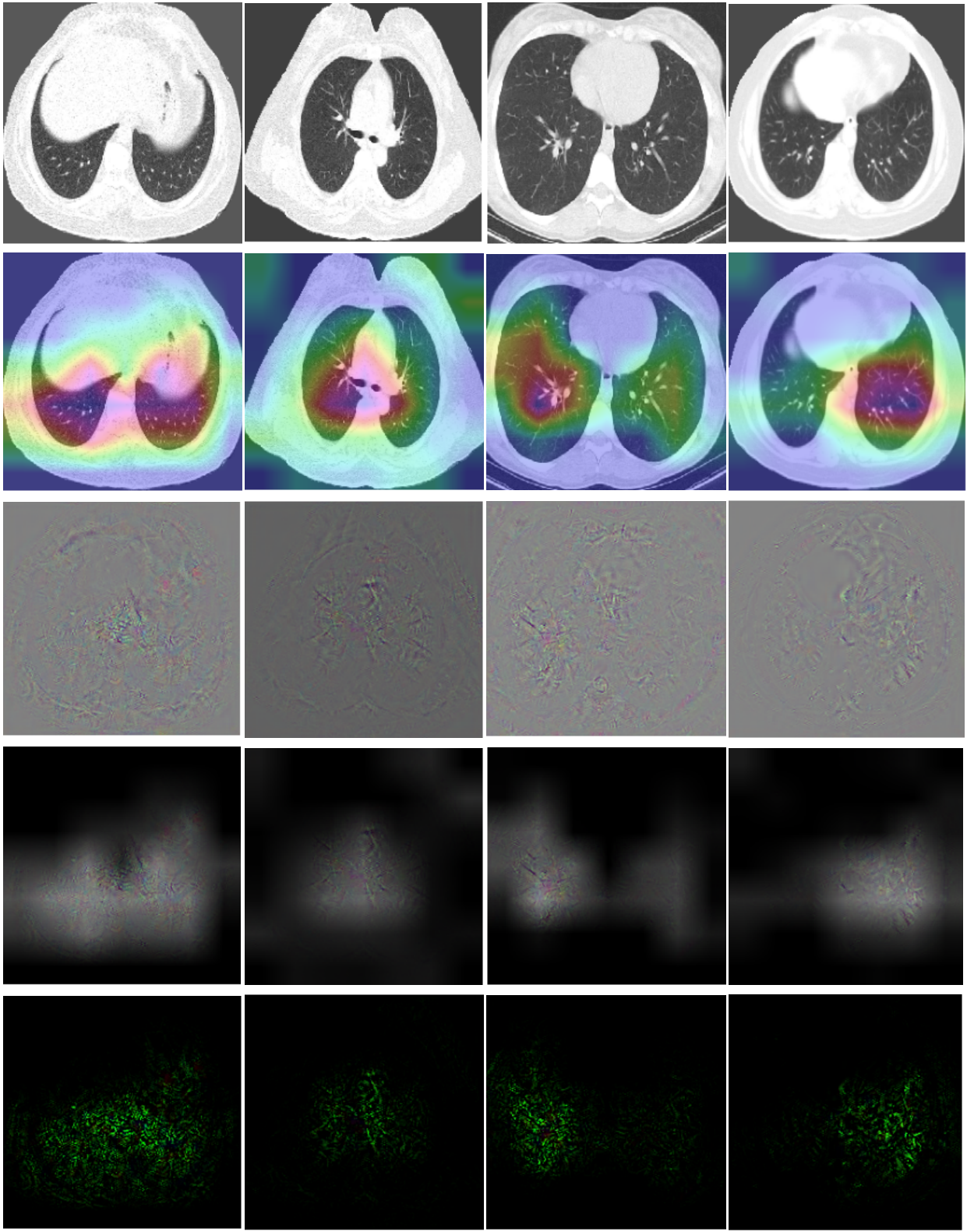

Figure 11: Sample chest CT images of normal cases and their associated characteristics identified by the proposed model.

## 4 Discussion

In this study, a deep transfer learning model based on the EfficientNetB7 [28] architecture was developed to detect COVID-19 from chest CT scans. While it's common to scale up ConvNets to achieve better accuracy, the EfficientNetB7 takes a unique approach that uniformly scales all three dimensions: depth, width, and image resolution. In contrast, the widely used ResNet can be scaled up from ResNet-18 to ResNet-200 by simply adding layers and thus depth. Additionally, most ConvNets put in practice are developed at a fixed resource cost and then scaled up to attain better accuracies. On the other hand, EfficientNet strived to create a more principled, simple, and efficient method to scale up a CNN by utilizing a simple but effective set of fixed compound coefficients. Furthermore, EfficientNetB7 has been proven to transfer well and achieve state-of-the-art accuracy on 5 out of 8 widely used datasets--all while reducing parameters by up to 21x than existing ConvNets [28]. In this study, the proposed EfficientNetB7 transfer learning model achieved a test accuracy of 92.71%.

To attain such an accuracy across a dataset as diverse and difficult to learn as the one employed in this paper, the fundamental issue in supervised machine learning known as overfitting [50] had to be surmounted. Four main methods of preventing overfitting were leveraged: data expansion, regularization, early stopping, and reduction of network complexity [50].

Initially, the dataset used in this study only consisted of images from CNCB [34], so overfitting was more likely to transpire due to lack of data diversity and high bias. Later, chest CT images were aggregated from TCIA [35] and COVID-CTset [36] to increase data diversity. As expected, data expansion made the greatest impact in improving accuracy.

Secondly, the exterior exclusion preprocessing algorithm served as a preventive measure against overfitting because of its regularization effect, which in turn helped with generalization. However, it was found that exterior exclusion had only improved accuracy, recall, and F1-score by 0.23%, 0.81%, and 0.15%, respectively, while decreasing the precision by 0.41%. This was especially surprising because contrary to popular belief, reducing noise had actually harmed part of the model's performance. It is hypothesized that because the dataset utilized in this study was already diverse enough and therefore had a low bias, applying a regularizing effect through a complicated augmentation was merely adding a touch of robustness to the model. The exterior exclusion augmentation was also programmed to apply to 50% of the images. The reason why it wasn't applied to all 17,698 images in the dataset is that it distorted some images and completely removed the lungs along with the background. Hence, 50% might have been too much and caused gradient descent to behave erratically, thus hindering convergence. This is supported by the jagged plots illustrated in Figures 4 and 5. However, it is still hard to say if exterior exclusion would have helped if a bias in the data was present, the number of images it was applied to was lower, or if it was a combination of the two.

Early stopping was also implemented to stop training when the validation accuracy stopped improving. Validation accuracy was chosen to be monitored over other metrics such as loss because it was the least erratic and best represented overall model performance. The patience, or the number of epochs to wait before early stopping halted training, was set to 15 to accommodate for the exterior exclusion preprocessing algorithm's regularization effect. Moreover, early stopping saved computational time.

Lastly, reducing the network complexity helped combat overfitting. The art of fine-tuning was a tedious process, but after experimentation with hyperparameters, it was found that a learning rate of 0.001 coupled with the Adam optimizer, a batch size of 32, image dimensions of 224 pixels by 224 pixels, and a fully connected layer with 512 neurons followed by a single-neuron fully connected layer presented the best results.

    A principal objective of this study was to maximize generalizability so that if deployed in the real world, the model would still accurately predict completely unseen images. From a technical standpoint, collecting more data is the safest and most reliable way to achieve such a goal. Hence, images were aggregated from the aforementioned three distinct sources. However, this made the training process a much more rigorous task because the model had to learn several types of images and pick up patterns that would otherwise be consistent in a dataset with images from a single source. Consequently, it was unsurprising to obtain a relatively lower classification accuracy of 92.71%. On the other hand, a considerably higher sensitivity of 95.79% was achieved--an attractive trait of deep learning-based diagnostics because of its improvement from RT-PCR testing methods. Since sensitivity is inversely proportional to the number of false negatives, a low number of COVID-19 cases were missed--a principal objective in this study. High sensitivity also ensures that in the real world, sick patients will rarely be predicted as healthy. If the sensitivity was low, COVID-19 positive patients would often be diagnosed as healthy, so they could be sent home and further the spread of COVID-19. Thus, the high sensitivity presents an encouraging result which is vital for the clinical management of the novel coronavirus.

    As portrayed in Table 1's model classification report, the precision of classifying COVID-19 chest CT scans is comparatively lower than its sensitivity. While this does mean that the model will have a fair number of false positives, potentially leading to unnecessary treatment for healthy patients, it further ensures that extremely few COVID-19 patients will be diagnosed as non-COVID-19 cases.

    In the interest of model transparency, a detailed visual analysis was utilized in the form of various Grad-CAM techniques. As observed in the qualitative results presented in Figures 8 and 9, respectively, the proposed model was capable of correctly detecting key features relevant to COVID-19 and normal chest CT scans. Altogether, leveraging explainability techniques opened the black box of deep learning.

    While impressive metrics have been attained, there are still a number of future steps to take and limitations to consider. Firstly, an area of concern lies within the data. Chest CT images indicative of common pneumonia should be included in the dataset so the model doesn't misclassify common pneumonia as COVID-19, which is possible because the two diseases contain similar features. The volume of data also needs to be increased, and although the current dataset has already been assembled from CNCB, TCIA, and COVID-CTset, supplementary data from additional sources would further prevent overfitting, improve robustness, and enhance generalizability. Other data preprocessing techniques could also be experimented with to reduce noise such as imaging artifacts or CT scanner beds. As a result, the model would base its decisions on more relevant indicators characteristic of COVID-19 such as consolidation, crazy-paving patterns, and ground-glass opacities. Additionally, different network architectures could be experimented with, as this study only considered the EfficientNetB7 pre-trained model for transfer learning. Next, a useful tool to understand how successive ConvNet layers transform their input is intermediate class activation maps. This would aid in visualizing the model's training process, so humans can validate the key patterns and features that are being learned. Subsequently, enhancing the current explainability algorithms to be more precise is a future step. Other popular visualization methods such as Layer-wise Relevance Propagation (LRP) or

Contrastive LRP could thus be leveraged. Severity assessments or patient risk stratification could even be created based on infection localizations and identified features. Once all these measures have been taken, real-world deployment of the deep learning framework on a web application could complement RT-PCR testing methods and serve as a second opinion to radiologists, alleviating the burden on the healthcare industry.

However, it is worth noting that although deep learning-based chest CT diagnostics present higher sensitivities in comparison to RT-PCR, screening tools should be reserved for extreme cases where potentially COVID-19 positive patients are in urgent need of an accelerated and more detailed form of diagnosis. Everyday testing should still defer to RT-PCR due to its ease of access and convenience.

## 5  Conclusion

This paper proposed a machine learning-based classification of COVID-19. A novel approach taken in assembling a diverse dataset and employing a wide variety of explainability techniques increased model generalizability and transparency. Visual analyses also provided insight into relevant features pertaining to COVID-19 and healthy lungs, which may benefit clinicians in CT lung screening. Finally, this study aims to encourage the continued research and development of COVID-19 in an effort to assist the healthcare industry and combat the global pandemic.

**Acknowledgements**

The author thanks Hayden Gunraj of the University of Waterloo for his continued guidance throughout this project.